\documentclass[]{aa}
\usepackage[varg]{txfonts}
\usepackage{graphicx}
\usepackage{amsmath}
\usepackage{amssymb}
\usepackage{footmisc}
\usepackage{color}
\usepackage{longtable}
\usepackage{natbib,twoopt}
\usepackage{mathabx}
\usepackage{graphics}
\usepackage{booktabs}
\usepackage{arydshln}

\newcommand{\OII}{[\mbox{O\,{\sc ii}}]}
\newcommand{\OIII}{[\mbox{O\,{\sc iii}}]}
\newcommand{\NII}{[\mbox{N\,{\sc ii}}]}

\begin{document} 

\title{Taking stock of SLSN and LGRB host galaxy comparison using a complete sample of LGRBs}

\author{
J. Japelj\inst{1}
\and
S. D. Vergani\inst{2,3} 
\and 
R. Salvaterra\inst{4}
\and 
L. K. Hunt\inst{5}
\and
F. Mannucci\inst{5}
}

\institute{INAF - Osservatorio Astronomico di Trieste, via G. B. Tiepolo 11, 34131 Trieste, Italy; \email{japelj@oats.inaf.it}
\and GEPI - Observatoire de Paris Meudon. 5 Place Jules Jannsen, F-92195, Meudon, France
\and INAF - Osservatorio Astronomico di Brera, via E. Bianchi 46, 23807 Merate, Italy
\and INAF - IASF Milano, via E. Bassini 15, I-20133, Milano, Italy
\and INAF - Osservatorio Astrofisico di Arcetri, Largo E. Fermi 5, I-50125 Firenze, Italy
}
 

\date{Received DD Mmmm YYYY / Accepted DD Mmmm YYYY} 
   
\abstract{Long gamma-ray bursts (LGRBs) and superluminous supernovae (SLSNe) are both explosive transients with very massive progenitor stars. Clues about the nature of the progenitors can be found by investigating environments in which such transients occur. While studies of LGRB host galaxies have a long history, dedicated observational campaigns have only recently resulted in a high enough number of photometrically and spectroscopically observed SLSN hosts to allow statistically significant analysis of their properties. In this paper we make a comparison of the host galaxies of hydrogen-poor (H-poor) SLSNe and the {\it Swift}/BAT6 sample of LGRBs. In contrast to previous studies we use a complete sample of LGRBs and we address a special attention to the comparison methodology and the selection of SLSN sample whose data have been compiled from the available literature. At intermediate redshifts ($0.3 < z < 0.7$) the two classes of transients select galaxies whose properties (stellar mass, luminosity, star-formation rate, specific star-formation rate and metallicity) do not differ on average significantly. Moreover, the host galaxies of both classes of objects follow the fundamental metallicity relation and the fundamental plane of metallicity. In contrast to previous studies we show that at intermediate redshifts the emission line equivalent widths of the two populations are essentially the same and that the previous claims regarding the  higher fraction of SLSN hosts among the extreme emission line galaxies with respect to LGRBs are mostly due to a larger fraction of strong-line emitters among SLSN hosts at $z < 0.3$, where samples of LGRB hosts are small and poorly defined.}
 
 
\authorrunning{} 
\titlerunning{ }
\maketitle
  
\section{Introduction}
Distinctly more luminous than ordinary supernovae, the recently established class of superluminous supernovae \citep[SLSN;][]{Quimby2011} is associated with the deaths of very massive stars \citep[e.g.][]{GalYam2009} and is being recognized as a new class of cosmic beacons that pinpoints distant star-forming galaxies \citep[][hereafter L14, L15, A16 and P16]{Lunnan2014,Leloudas2015,Angus2016,Perley2016d} and lights up their environment \citep{Berger2012,Vreeswijk2014}. Due to their extreme luminosities, together with peculiar spectral and light curve properties, understanding the nature of the progenitors and emission mechanisms involved in SLSNe has proved to be rather challenging.

SLSNe come in many flavours, which can be grouped into two subclasses based on the presence of hydrogen in their spectra \citep{GalYam2012}: hydrogen-rich SLSN-II and hydrogen-poor (H-poor) SLSN-I. The extreme luminosities of the former are commonly explained as an interaction of supernova ejecta with a dense interstellar medium \citep[e.g.][but see \citealt{Inserra2016}]{Moriya2013}. The physical nature of the SLSN-I type however remains debated. A subclass of slowly evolving SLSN-I was proposed to be powered by radioactive decay of \element[][56]{Ni} \citep[][]{GalYam2009,GalYam2012}. The amount of \element[][56]{Ni} required in this case should be large, but could be produced by a pair-instability SN \citep{Barkat1967}. Such SN is expected to arise in the case of a very massive star ($\sim 140-260$ M$_{\odot}$), in which the cascading conversion of star-supporting photons to electron-positron pairs is followed by a rapid contraction and thermonuclear explosion. On the other hand, SLSN-I emission could be powered by a central engine through the reheating of the SN ejecta by an accretion onto a central black hole \citep{Dexter2013} or spin-down of a rapidly-rotating neutron star with strong magnetic fields (i.e. magnetar model; \citealt{Woosley2010,Dessart2012,Inserra2013,Nicholl2013,Nicholl2015}).

In addition to studying SLSN light curves and spectra, clues about the progenitors of SLSN can be found by investigating the environments in which they occur. SLSN host galaxies are often characterized by strong nebular emission lines (L14; L15; P16). In cases where the host is experiencing a recent and young starburst, the equivalent widths of the emission lines can give a meaningful constraint on the age of the stellar population and therefore the age and mass of a progenitor star \citep{Thoene2015}. Properties of the environment can also provide clues to distinguish between more or less probable progenitor models. For example, progenitor stars are expected to be of low metallicity in order to produce the rapidly rotating neutron star \citep{Woosley2006,Yoon2006} or pair-instability SN \citep{Langer2007} and therefore their host environment is expected to be metal-poor. Studies performed so far have shown that the H-poor SLSNe are typically found in host galaxies of low luminosity, low stellar mass (M$_{\star}$), low star formation rates (SFRs) and low nebular metallicities \citep[][L14; L15; A16; P16]{Neill2011}. On the other hand, host galaxies of SLSN-II type are found in galaxies spanning a larger range in luminosity, mass and SFR.

\begin{table*}
\small
\renewcommand{\arraystretch}{1.3}
\begin{center}
\begin{tabular}{lccccccr}
\hline
\hline
SLSN & redshift & M$_{\rm B}$ & $\log$ M$_{\star}$ & SFR & 12 + $\log \left( \frac{\rm O}{\rm H}\right)$  & References\\
        &              &  (mag)           & (M$_{\odot}$)         & (M$_{\odot}$yr$^{-1}$) & M08                                 & \\
\hline
MLS12110  & 0.303 & -19.39 & 9.25$_{-0.06}^{+0.04}$ & 0.73 $\pm$ 0.02 & 8.54$_{-0.05}^{+0.05}$ & 1, 2\\
PTF12mxx & 0.33   & -16.9    & 8.16$_{-0.24}^{+0.61}$ & 0.11$_{-0.02}^{0.04}$ & 8.69$_{-0.46}^{+0.23}$& 4, 4\\
PTF09cwl   & 0.349 & -15.7 & 7.85$_{-0.33}^{+0.23}$& $<0.06$                   & -                                    & 4,4\\
PTF10bjp   & 0.359 & -18.8   & 9.21$_{-0.17}^{+0.15}$ & 0.93$_{-0.25}^{+0.28}$ & 8.24$_{-0.23}^{+0.13}$ & 4, 4\\
SN2006oz  & 0.396 & -16.96 & 8.67$_{-0.04}^{+0.11}$ & 0.13 $\pm$ 0.11 & 8.46$_{-0.13}^{+0.08}$ &  1, 3\\
PTF10vqv   & 0.45   & -18.4   & 8.08$_{-0.15}^{+0.92}$ &1.36$_{-0.36}^{+0.41}$ & 8.51$_{-0.08}^{+0.06}$ & 4, 4\\
PTF09atu   & 0.501 & -16.3& 8.35$_{-0.62}^{+0.33}$& $<0.11$                    & -                                    & 4, 4\\
PS1-12bqf   & 0.522 & -20.28 & 9.45$_{-0.10}^{+0.13}$ & 1.05 $\pm$ 0.55 & 8.82$_{-0.18}^{+0.20}$ &  2, 2\\
PS1-11ap    & 0.524 & -18.83 & 8.48$_{-0.12}^{+0.15}$ & 0.30 $\pm$ 0.18 & 8.42$_{-0.12}^{+0.10}$ & 2, 2\\
PS1-10bzj   & 0.650 & -17.90 & 7.21$_{-0.05}^{+0.11}$ & 5.85 $\pm$ 2.10 & 7.76$_{-0.12}^{+0.15}$ & 2, 2\\
PS1-12zn    & 0.674 & -18.75 & 8.34$_{-0.32}^{+0.24}$ & 4.50 $\pm$ 2.00 & 8.40$_{-0.07}^{+0.07}$ & 2, 2\\
\hline
\hline
\end{tabular}
\end{center}
\caption{Properties of H-poor SLSN sample used in the paper for cumulative plots. All objects have available near-infrared photometric observations and therefore reliably measured stellar masses. Reported are SLSN designation, redshift, absolute {\it B}-band magnitude, stellar mass, star formation rate and nebular metallicity (in the M08 metallicity calibration). The last column reports the references from which the stellar masses and spectra were obtained, respectively. Masses and $M_{\rm B}$ were adopted directly from the referenced works, while SFR and metallicities were measured as described in Section 2.\newline References: (1) \citet{Angus2016} (2) \citet{Lunnan2014} (3) \citet{Leloudas2015} (4) \citet{Perley2016d}.
}
\label{tab1}
\end{table*}

To put the results into perspective, SLSN hosts have been compared to the host galaxies of other types of explosive transients with massive progenitors. Of particular interest is a comparison with long duration gamma-ray bursts \citep[LGRBs; e.g.][]{Kumar2015}, whose optical afterglow luminosity shortly after the explosion can vastly exceed that of SLSNe \citep[e.g.][]{Bloom2009}, and for which the clues of their progenitors are likewise being sought in their host galaxy environment \citep[e.g.][]{LeFloch2003,Savaglio2009,Levesque2010,Graham2013,Kruhler2015,Vergani2015,Perley2015b,Japelj2016}. In addition to the collapsar scenario \citep[e.g.][]{Woosley1993} the magnetar has also been invoked as a possible central engine of LGRBs and their accompanying supernovae \citep{Usov1992,Metzger2015,Greiner2015,Cano2016}. If the central engine for the two transients is indeed the same, it is expected that SLSN and LGRB are preferentially found in the same environments (though the opposite is not necessarily true). Comparing the host galaxies of SLSN-I and LGRBs, L14 finds them to be similar in M$_{\star}$, SFR, specific SFR (SFR divided by stellar mass) and metallicity, especially if excluding the events at the lowest redshifts. On the contrary, spectroscopic study of L15 and photometric study of A16 conclude that the LGRB hosts are found in less extreme environments (e.g. higher M$_{\star}$ and SFRs and lower emission line EWs) than SLSN-I. 

The reason for different conclusions could at least partly be attributed to the methodology and biased samples of both SLSNe and LGRBs used in the comparison. We therefore aim to take stock of this issue by comparing the properties of the host galaxies of the {\it Swift}/BAT6 complete sample of LGRBs \citep{Salvaterra2012} and a sample of SLSN host galaxies carefully selected from the literature. We discuss the effects the selection of the SLSN sample has on our conclusions and emphasise the importance of a redshift interval assumed in the comparison of the two galaxy populations. Because of the limited statistics for the H-rich class of SLSNe (SLSN-II), our discussion is limited to the host galaxies of H-poor SLSNe. 

All errors are reported at 1$\sigma$ confidence unless stated otherwise. We use a standard cosmology \citep{Planck2014}: $\Omega_{\rm m} = 0.315$, $\Omega_{\Lambda} = 0.685$, and $H_{0} = 67.3$ km s$^{-1}$ Mpc$^{-1}$. All quantities are computed with respect to the Chabrier initial mass function \citep{Chabrier2003}.

\begin{figure*}[]
\centering
\includegraphics[scale=0.58]{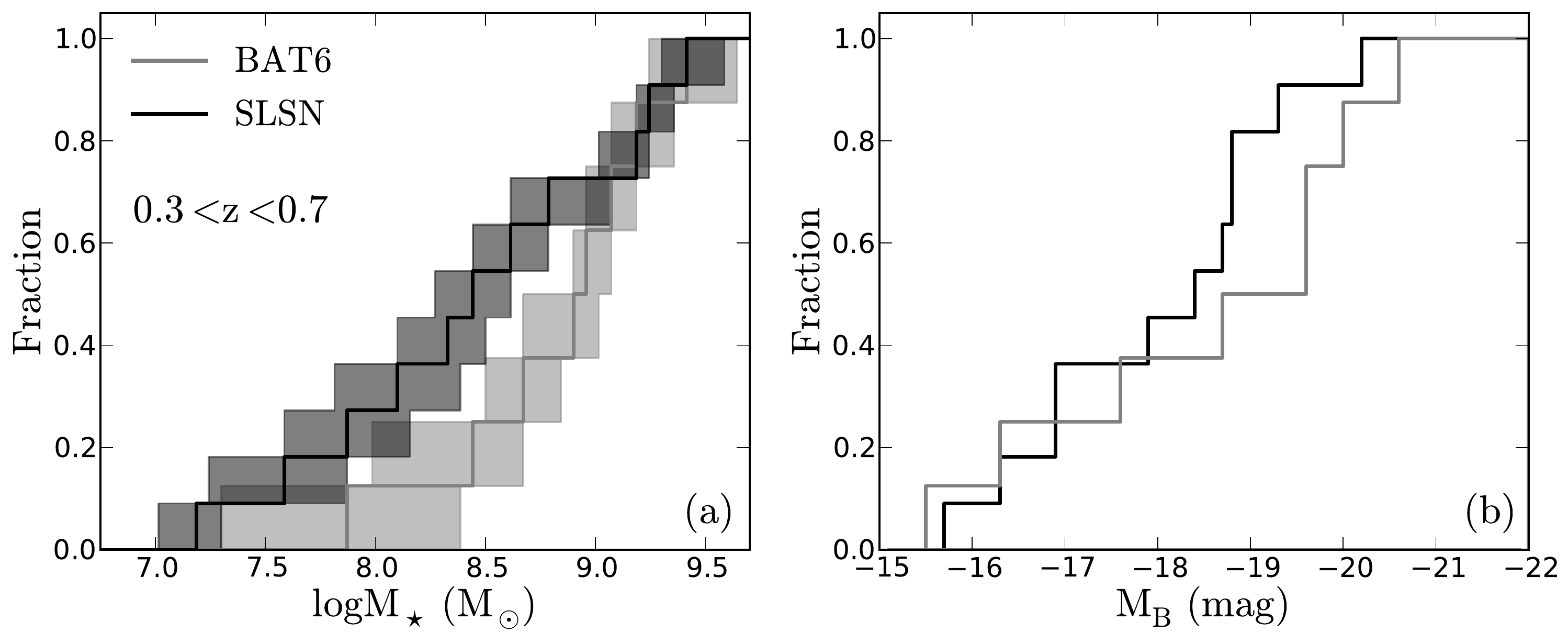}
\caption{Cumulative distributions of {\it (a)} stellar mass and {\it (b)} absolute {\it B}-band magnitude of SLSN and LGRB samples in the $0.3 < z < 0.7$ redshift range. For stellar masses the distributions were obtained via MC simulation taking into account the errors of individual measurements.}
\label{fig1}
\end{figure*}

\begin{figure*}[]
\centering
\includegraphics[width=\textwidth]{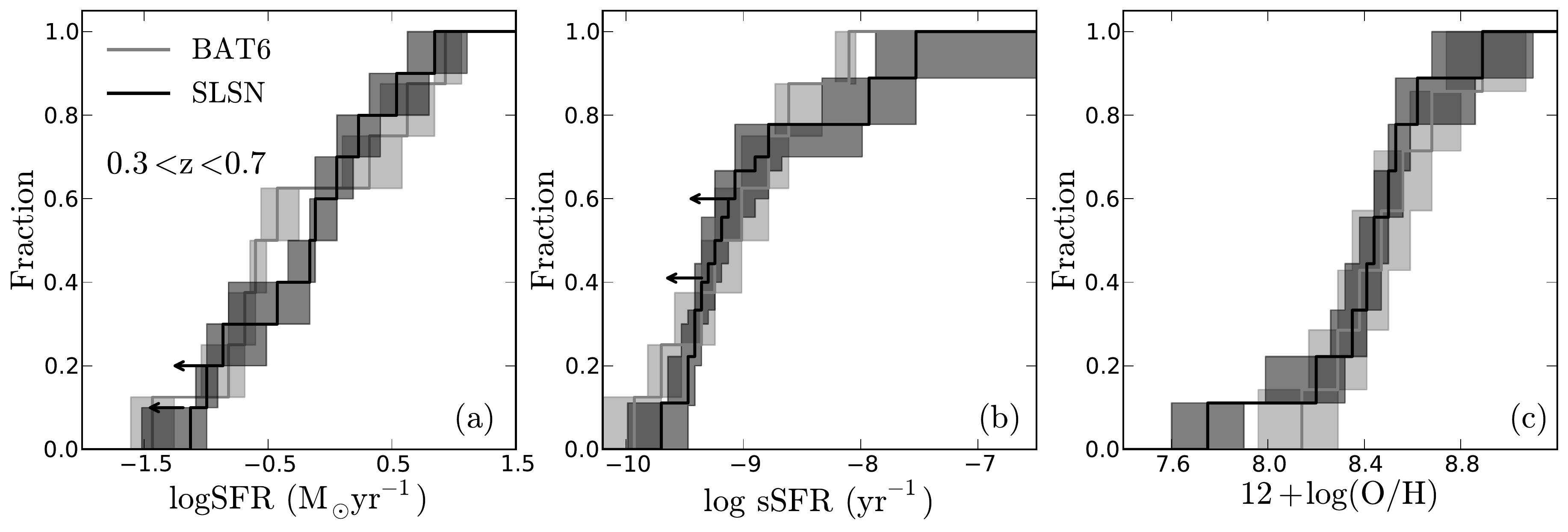}
\caption{Cumulative distributions of {\it (a)} star formation rate,  {\it (b)} specific SFR and {\it (c)} metallicities of SLSN and LGRB samples in the $0.3 < z < 0.7$ redshift range.}
\label{fig2}
\end{figure*}

\section{Data}
\label{selection}

As LGRB host galaxy sample we used the one based on the complete {\it Swift}/BAT6 sample of LGRBs \citep{Salvaterra2012}. The selection of this sample of LGRBs is based basically on the prompt emission only. In principle this does not introduce any bias on the host galaxy properties. Indeed, no correlations have been found to date between the prompt emission properties and those of the LGRB host galaxies \citep{Levesque2010b, Japelj2016}. The stellar masses, absolute $M_{\rm B}$ magnitudes, SFR and metallicity of the host galaxies are published for the $0.3<z<1.0$\footnote{There is also an object (GRB\,060614) at $z<0.3$. As its nature is debated, we exclude it from this analysis.} part of the sample \citep{Vergani2015,Japelj2016}. 

From the SLSN side, we focus our study on hydrogen-poor SLSNe, in which we include both the SLSN-I and SLSN-R type events, in the same redshift range of the GRB sample ($0.3<z<1.0$). After a close inspection of the data sets from the samples of L14, L15, A16 and P16, we noticed that the stellar masses of the overlapping subsamples differ, in some cases by an order of magnitude (where necessary, we transformed the masses to the Chabrier IMF calibration for a valid comparison). Stellar masses in all four works were obtained by modelling the photometric spectral energy distributions (SEDs) of the galaxies with synthetic models. Such high discrepancy is worrisome and therefore our selection of the sample was based foremost on the criterion of the quality of stellar mass determination. We therefore discard the stellar masses from L15 as their photometry and SED analysis is only referenced to a yet unpublished work, as well as all the stellar masses determined from SED without a rest-frame NIR observations. Indeed, we noticed a better agreement for the cases in which the stellar masses of the different works in the literature were determined from SED including NIR observations. It turned out that all published hosts in a redshift interval $0.3<z<0.7$ have available NIR observations, therefore we decided to use this redshift interval (both for LGRB and SLSN samples) for the comparison, to avoid introducing any additional bias due to the requirement of NIR observations when comparing average galaxy properties in Section 3.1. Our SLSN host galaxy sample is comprised of all the objects (11) in the redshift interval $0.3<z<0.7$ presented in L14, A16 and P16. The sample is summarized in Table 1. Based on the availability, stellar masses were adopted from the literature in the following order of preference reflecting the quality of the SEDs: P16, A16, and L14. Spectroscopic observations are available for all the galaxies in the sample.

To measure SFRs and metallicities of the part of SLSN host galaxies coming from L14 and A16 only, we used published flux densities of detected host emission lines and analyzed them by following the same prescriptions as used by \citet{Japelj2016} for the LGRB host galaxies. SFRs were measured from extinction-corrected spectral emission lines. The lines we used as SFR tracers were (in the order of preference and based on the availability): H$\alpha$, H$\beta$ \citep{Kennicutt1998} and \OII\, \citep{Kruhler2015}. Metallicities were computed by minimising all metallicity calibrations available for an emission-line set of each host galaxy \citep{Maiolino2008}. Fort the part of SLSN host galaxies coming from P16, we used the values reported in the paper as they comply to the same prescriptions used for the LGRB host galaxy spectra. Absolute {\it B}-band magnitudes of the SLSN hosts were taken from L14. For the P16 hosts, $M_{\rm B}$ were either extrapolated from their SEDs or from the available photometry in their Table 2.

\begin{table*}
\small
\renewcommand{\arraystretch}{1.3}
\begin{center}
\begin{tabular}{lcccccr}
\hline
\hline
SLSN & redshift & $\log$ M$_{\star}$ & SFR & 12 + $\log \left( \frac{\rm O}{\rm H}\right)$  & References\\
        &  (M$_{\odot}$)         & (M$_{\odot}$yr$^{-1}$) & M08                                 & \\
\hline
PTF11hrq  & 0.057 & 8.55$_{-0.21}^{+0.28}$ & 0.20$_{-0.04}^{+0.04}$ & 8.28 $_{-0.03}^{+0.04}$ & 4, 4\\
PTF10hgi   & 0.098 & 7.95$_{-0.19}^{+0.18}$ & 0.04 $\pm$ 0.04 & 8.62$_{-0.10}^{+0.08}$ & 4, 3\\
PTF12dam & 0.1078 & 8.30$_{-0.15}^{+0.15}$ & 4.78$_{-1.17}^{+0.96}$& 8.08$_{-0.03}^{+0.02}$ & 4, 4\\
PTF10nmn & 0.123 & 8.79$_{-0.15}^{+0.11}$ & 0.53$_{-0.13}^{+0.12}$& 8.27$_{-0.09}^{+0.07}$ & 4, 4\\
SN1999as  & 0.127 & 9.34$_{-0.00}^{+0.08}$ & 0.17 $\pm$ 0.19 & 8.78$_{-0.12}^{+0.12}$ & 1,3\\
SN2007bi   & 0.127 & 8.13$_{-0.17}^{+0.31}$ & 0.02 $\pm$ 0.01 & 8.36$_{-0.30}^{+0.16}$ &  1,3\\
SN2011ke   & 0.143 & 6.90$_{-0.15}^{+0.17}$ & 0.40 $\pm$ 0.03 & 7.86$_{-0.08}^{+0.08}$ & 4, 3\\
PTF10bfz    & 0.175 & 8.65$_{-0.32}^{+0.11}$ & 0.39$_{-0.10}^{+0.10}$& 8.02$_{-0.12}^{+0.09}$& 4, 4\\
SN2012il    & 0.169 & 8.45$_{-0.17}^{+0.27}$ & 0.38 $\pm$ 0.07 & 7.86$_{-0.10}^{+0.08}$ & 1, 3\\
PTF10vwg  & 0.19 & 8.25$_{-0.59}^{+0.18}$ & $<$ 0.07& - & 4, 4\\
PTF11rks    & 0.192 & 9.11$_{-0.16}^{+0.13}$ & 0.29 $\pm$ 0.13 & 8.70$_{-0.32}^{+0.20}$ &  4, 3\\
PTF10aagc  & 0.206 &8.98$_{-0.21}^{+0.13}$ & 0.47$_{-0.16}^{+0.19}$& 8.33$_{-0.12}^{+0.08}$&  4,4\\
SN2010gx   & 0.230 & 7.87$_{-0.21}^{+0.13}$ & 0.42 $\pm$ 0.17 & 7.78$_{-0.14}^{+0.12}$ & 4, 3\\
SN2011kf    & 0.245 & 8.09$_{-0.32}^{+0.27}$ & 0.15 $\pm$ 0.05 & 8.14$_{-0.14}^{+0.12}$ &  1, 3\\
PTF09cnd    & 0.258 & 8.32$_{-0.18}^{+0.15}$ & 0.23 $\pm$ 0.06 & 8.44$_{-0.30}^{+0.14}$ &  4, 3\\
SN2005ap   & 0.283 & 8.46$_{-0.15}^{+0.16}$ & -                          & -                                    & 1,-\\
PTF10uhf$^{(*)}$ & 0.289 & 10.6$_{-0.2}^{+0.2}$     & 5$_{-2}^{+2}$       & 8.98$_{-0.01}^{+0.01}$ & 4,4\\
MLS12110   & 0.303 & 9.25$_{-0.06}^{+0.04}$ & 0.73 $\pm$ 0.02 & 8.54$_{-0.05}^{+0.05}$ & 1, 2\\
PTF12mxx   & 0.33 & 8.16$_{-0.24}^{+0.61}$ & 0.11$_{-0.02}^{0.04}$ & 8.69$_{-0.46}^{+0.23}$& 4, 4\\
PTF09cwl    & 0.349 &7.85$_{-0.33}^{+0.23}$& $<0.06$                   & -                                    & 4,4\\
PTF10bjp    & 0.359 & 9.21$_{-0.17}^{+0.15}$ & 0.93$_{-0.25}^{+0.28}$ & 8.24$_{-0.23}^{+0.13}$ & 4, 4\\
SN2006oz  & 0.396 & 8.67$_{-0.04}^{+0.11}$ & 0.13 $\pm$ 0.11 & 8.46$_{-0.13}^{+0.08}$ &  1, 3\\
PTF10vqv   & 0.45 & 8.08$_{-0.15}^{+0.92}$ &1.36$_{-0.36}^{+0.41}$ & 8.51$_{-0.08}^{+0.06}$ & 4, 4\\
PTF09atu   & 0.501 & 8.35$_{-0.62}^{+0.33}$& $<0.11$                    & -                                    & 4, 4\\
PS1-12bqf   & 0.522 & 9.45$_{-0.10}^{+0.13}$ & 1.05 $\pm$ 0.55 & 8.82$_{-0.18}^{+0.20}$ &  2, 2\\
PS1-11ap     & 0.524 & 8.48$_{-0.12}^{+0.15}$ & 0.30 $\pm$ 0.18 & 8.42$_{-0.12}^{+0.10}$ & 2, 2\\
PS1-10bzj    & 0.650 & 7.21$_{-0.05}^{+0.11}$ & 5.85 $\pm$ 2.10 & 7.76$_{-0.12}^{+0.15}$ & 2, 2\\
PS1-12zn     & 0.674 & 8.34$_{-0.32}^{+0.24}$ & 4.50 $\pm$ 2.00 & 8.40$_{-0.07}^{+0.07}$ & 2, 2\\
\hline
\hline
\end{tabular}
\end{center}
\caption{Properties of SLSN sample used to plot the relations in Figure \ref{fig3}. The table lists all SLSN hosts complying with our selection criteria (Section \ref{selection}), even though only hosts with detected emission lines and measured metallicities are used for the plotting. Reported are SLSN designation, redshift, stellar mass, star formation rate, nebular metallicity (in the M08 metallicity calibration). The last column reports the references from which the stellar masses and spectra were obtained, respectively. Masses and $M_{\rm B}$ were adopted directly from the referenced works, while SFR and metallicities were measured as described in Section 2.\newline References: (1) \citet{Angus2016} (2) \citet{Lunnan2014} (3) \citet{Leloudas2015} (4) \citet{Perley2016d}.\newline * Mass and SFR scaled for a factor of 4 to get the approximate values at the SLSN site, as suggested by \citep{Perley2016d}.
}
\label{tab2}
\end{table*}

\section{Results}

\subsection{Cumulative properties in the 0.3 < z < 0.7 range}
For the comparison of most of the host galaxy properties it is fundamental to restrict the analysis to similar redshift ranges, in order to avoid introducing systematic differences due to galaxy evolution with redshift. Therefore, for the cumulative distributions we consider only a redshift range common to our LGRB and SLSN sample of host galaxies, i.e. $0.3<z<0.7$. Mean redshifts of the SLSN (11 hosts) and LGRB (8 hosts) samples are $<z> = 0.46$ and 0.53, respectively.

In Figure \ref{fig1} we compare cumulative distributions of stellar masses and absolute magnitudes of the two populations. Both the number of SLSN and LGRB hosts in the samples is small, therefore we took the errors into account (by performing MC simulation), which is shown by shaded regions in the plots (except for the $M_{\it B}$ values which in the literature are reported without errors). According to the Kolmogorov-Smirnov test, the probability that the two samples are drawn from the same distribution is $p_{\rm KS} \sim 0.15$ and $p_{\rm KS} \sim 0.32$ for masses and luminosities, respectively. In both cases we cannot discard the hypothesis that the two samples are drawn from the same distribution.

The SLSN host galaxy sample can be affected by selection biases (see the relevant discussion in L14 and P16). There are two factors affecting the selection of SLSN that can result in a bias of the host galaxies against dusty, massive galaxies. First, the SLSN are optically selected and therefore more difficult to be detected in dusty galaxies. Second, the spectroscopic follow-up of SLSN candidates is preferentially carried out for the events significantly brighter than their hosts (see L14). To verify the robustness of our findings, we therefore artificially added massive hosts of $\log {\rm M}_{\star}/{\rm M}_{\odot} = 11$ to the SLSN sample and we determined how many of them are needed to discard the hypothesis that the GRB and SLSN host populations are similar. 
We find that to reach that result (i.e. $p_{\rm KS} \lesssim 0.01$) we should add $\sim 15$ massive galaxies to the sample.
It is hard to believe that more than half of the SLSN would be lost due to their origin in such massive galaxies (see also discussion in P16), therefore we conclude that this does not affect our conclusion. 
The conditions we are imposing to select our SLSN sample from those published in the literature do not introduce any further bias. In any case, we repeated the test also by adding low-mass ($\log {\rm M}_{\star}/{\rm M}_{\odot} = 7$) hosts, and we find that the SLSN distribution differed significantly from the one of LGRBs when $\sim 10$ low-mass host galaxies are added to the sample. 
Our results on the stellar masses and luminosities, therefore, do not support the hypothesis that SLSN and GRB host galaxies belong to two different populations. Finally, the uncertainty regions in the plots take only measurement errors into account. To understand the impact of small sample sizes on our conclusions, we estimate the uncertainty region by performing a bootstrapping resampling-with-replacement  MC analysis. The results give similar conclusions as before: we cannot discard the hypothesis that the two samples are drawn from the same distribution.

In Fig.\,2 we report the plots resulting from the investigation of the SFR, specific SFR (sSFR) and metallicities of SLSN and LGRB hosts. The distributions of these properties are found to be very similar at this redshift range for both classes of objects. Two SLSN hosts lack a significant detection of emission lines in their spectra. Performing a similar test as in the case of stellar masses (e.g. by adding artificial values to the distributions and looking for the point when the discrepancy between distributions becomes significant) it is clear that an unlikely high portion of a SLSN population would have to be missed in the surveys in order to make the SLSN host properties different from those of LGRB hosts.

\begin{figure*}[t]
\centering
\includegraphics[width=\textwidth]{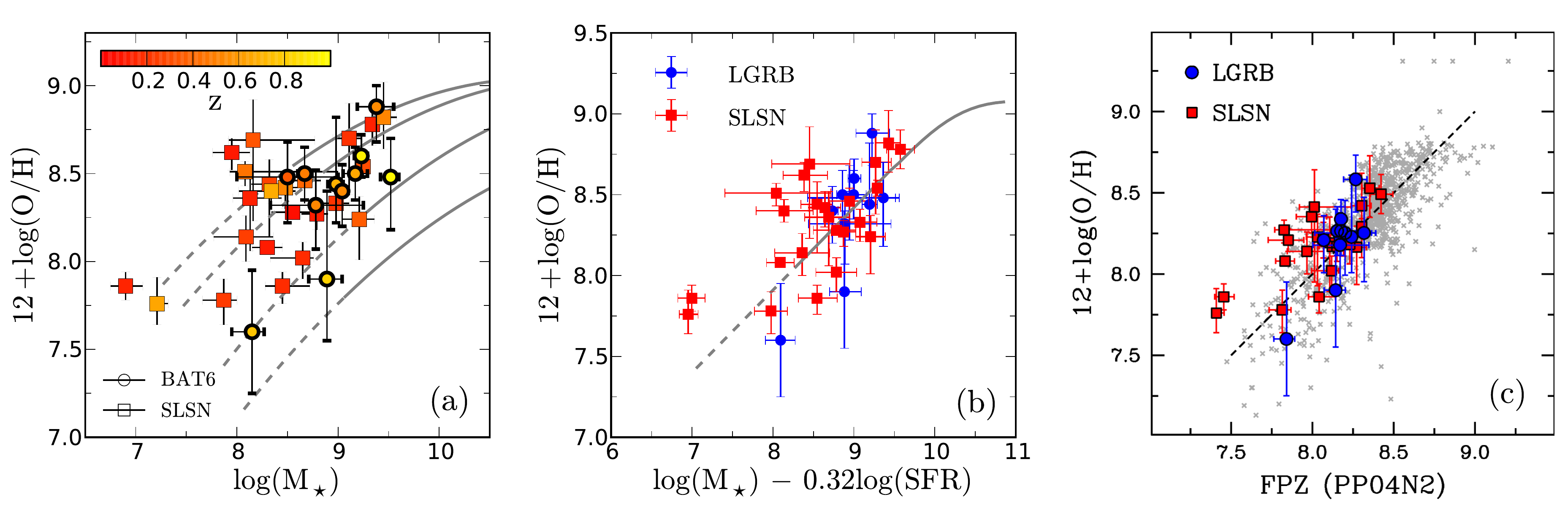}
\caption{Comparing the full LGRB and SLSN samples to relations that are followed by field star-forming galaxy populations. {\it (a)} Mass-metallicity relation. Overplotted are the median relations of star-forming galaxies at redshifts (from top to bottom) $z =$ 0.07, 0.7, 2.2 and 3.5 \citep{Mannucci2009}. Data are color-coded according to their redshift. {\it (b)} Fundamental metallicity relation \citep[line;][]{Mannucci2010,Mannucci2011} for SLSNe and LGRBs. {\it (c)} SLSN and LGRB host galaxies in the fundamental metallicity plane (FPZ) for low-mass galaxies \citep[grey;][]{Hunt2016}. The plane plotted is done with \NII/H$\alpha$ metallicity calibration \citep[][]{Pettini2004}. Dashed lines in {\it (a)} and {\it (b)} show the extrapolations of the relations towards lower masses, where the relations are poorly defined.}
\label{fig3}
\end{figure*}

\subsection{Relations between M$_{\star}$, SFR and metallicity: comparison to star-forming galaxies}
We then check the behaviour of SLSN host galaxies with respect to the relations found between $M_{\star}$, SFR and metallicity (mass-metalicity, FMR \citealt{Mannucci2010}, and FPZ \citealt{Hunt2016}). Apart from the mass-metalicity relation \citep{Tremonti2004,Mannucci2009}, these relations are thought to be insensitive to redshift (at least at the redshift considered in this work). Here we therefore include all the LGRB hosts of the BAT6 $0.3 < z < 1$ sample, and we expand the SLSN sample by including the low-redshift ($z < 0.3$) SLSN hosts. This extended SLSN host galaxy sample includes all the objects (30) at z<0.7 studied by L14, A16 and P16, with exception of two objects that do not have NIR observations (see Table 2 for the full account of the SLSN sample). We then add the requirement of the determination of the metallicity. At the end, our samples are made of 11 LGRB hosts (over the 13 objects of the BAT6 sample at $0.3 < z < 1$; see also \citealt{Japelj2016}) and of 24 SLSN hosts (over 30 objects of the union of the L14, A16 and P16 samples at $z<0.7$).

In Fig. \ref{fig3}a we plot the mass vs metallicity of the two samples at $z<1$. Focusing on the $0.3<z<0.7$ range, it seems that SLSN and LGRB hosts occupy the same region in the mass-metallicity relation and also agree on average with the median relations of field star-forming galaxies at the same redshifts. Approximately half of the low-redshift  ($z < 0.3$) SLSN hosts are found to have significantly lower metallicities as the star-forming galaxies at the same stellar mass and redshift range.

LGRB host galaxies have been found to follow the fundamental metallicity relation (FMR), a two-dimensional surface defined by metallicities, SFRs and stellar masses of the main-sequence star forming galaxies \citep{Mannucci2011,Japelj2016}. In Figure \ref{fig3} we show that SLSN host galaxies also lie in the same plane with a similar dispersion as LGRB hosts: the dispersion of LGRB and SLSN samples around the FMR relation is 0.22 dex and 0.26 dex, respectively. These values are larger than the typical dispersion of the FMR at $\log (M_{\star}/M_{\odot}) \gtrsim 9.4$ which can be attributed to small number statistics and increasing intrinsic scatter of the FMR relation towards lower values of $\mu_{0.32}=\log M_{\star} - 0.32\log SFR$ \citep{Mannucci2011}. Both LGRB and SLSN hosts are predominantly low-mass galaxies with masses extending to the range at which the FMR is not well defined. \citet{Hunt2012,Hunt2016} computed a scaling relation using stellar masses, SFRs and metallicities toward low-mass starburst galaxies to build a fundamental metallicity plane (FPZ). We plot our LGRB and SLSN host galaxy samples in the FPZ and show that they lie within the region populated by other types of low-mass galaxies, with dispersion of 0.16 dex and 0.18 dex, respectively.

\section{Discussion}
Previous studies of SLSN host galaxies, from which our SLSN data have been compiled, reached different conclusions regarding the relative properties of SLSN and LGRB host galaxies. In order to get a more robust understanding of the two populations of host galaxies: {\it (i)} we used a complete sample of LGRBs for the comparison; {\it (ii)} we require the NIR photometry for the stellar mass determination; {\it (iii)} we limited our analysis in the relevant cases to a narrower redshift range so as to minimize the impact of the evolution of the galaxy properties. Some of our results differ from the conclusions of previous studies.

L14 and L16 have found that, on average, SLSN hosts are much fainter than LGRB hosts. Because the faintness implies low masses and low metallicities, the result was inconsistent with the observed similarity between metallicity distributions of the two classes. On the contrary, we find that $M_{\rm B}$ and stellar mass distributions of the two classes are similar. The (statistically not significant) discrepancies of stellar masses could be due to the bias of SLSN sample against dusty/luminous host galaxies (see L14, P16). We note that even if the slight preference for lower masses is intrinsic to the  SLSN population, the effect is much smaller than what is claimed by some of the previous works. The similar nature of the LGRB and SLSN hosts is further corroborated by resolved imaging of the hosts. Using HST imaging, \citet{Lunnan2015} showed that SLSN and LGRB hosts have similar morphologies, effective galaxy radii and SFR densities. They also found that SLSN are tracing UV-bright regions in their hosts, though the correspondence is not as strong as in the case of LGRBs \citep{Fruchter2006}.

SLSN hosts are notable for their strong emission lines. L15 compared the equivalent width distribution of  the \OIII$\lambda5007$ line for the LGRB and SLSN hosts and claimed that the fraction of extreme emission line galaxies (EELGs, e.g. \citealt{Atek2011}) with high emission line equivalent widths is much higher among SLSN hosts with respect to the LGRB hosts. Because the strength of the EW$_{5007}$ diminishes with time after a starburst \citep[e.g.][]{Zackrisson2001,Inoue2011}, L15 concluded that the SLSNe represent host galaxies viewed shortly after a starburst event. Moreover, taking at face value the results by \citet{Zackrisson2001}, the EW$_{5007}$ is expected to strongly decrease after a few Myr corresponding roughly to the ZAMS lifetime of stars with masses of 50-60 M$_\odot$ \citep{Meynet2005}. Therefore, assuming that both SLSN and LGRB progenitor are formed during an (instantaneous) starburst event, SLSN that originated by more massive, short-living progenitors would be detected in galaxies with enhanced EW$_{5007}$, while LGRBs would be related to the death of 20-30 M$_\odot$ \citep[e.g.][]{Thoene2008,Levesque2010} and found in less extreme conditions. 

\begin{figure}[t]
\centering
\includegraphics[scale=0.42]{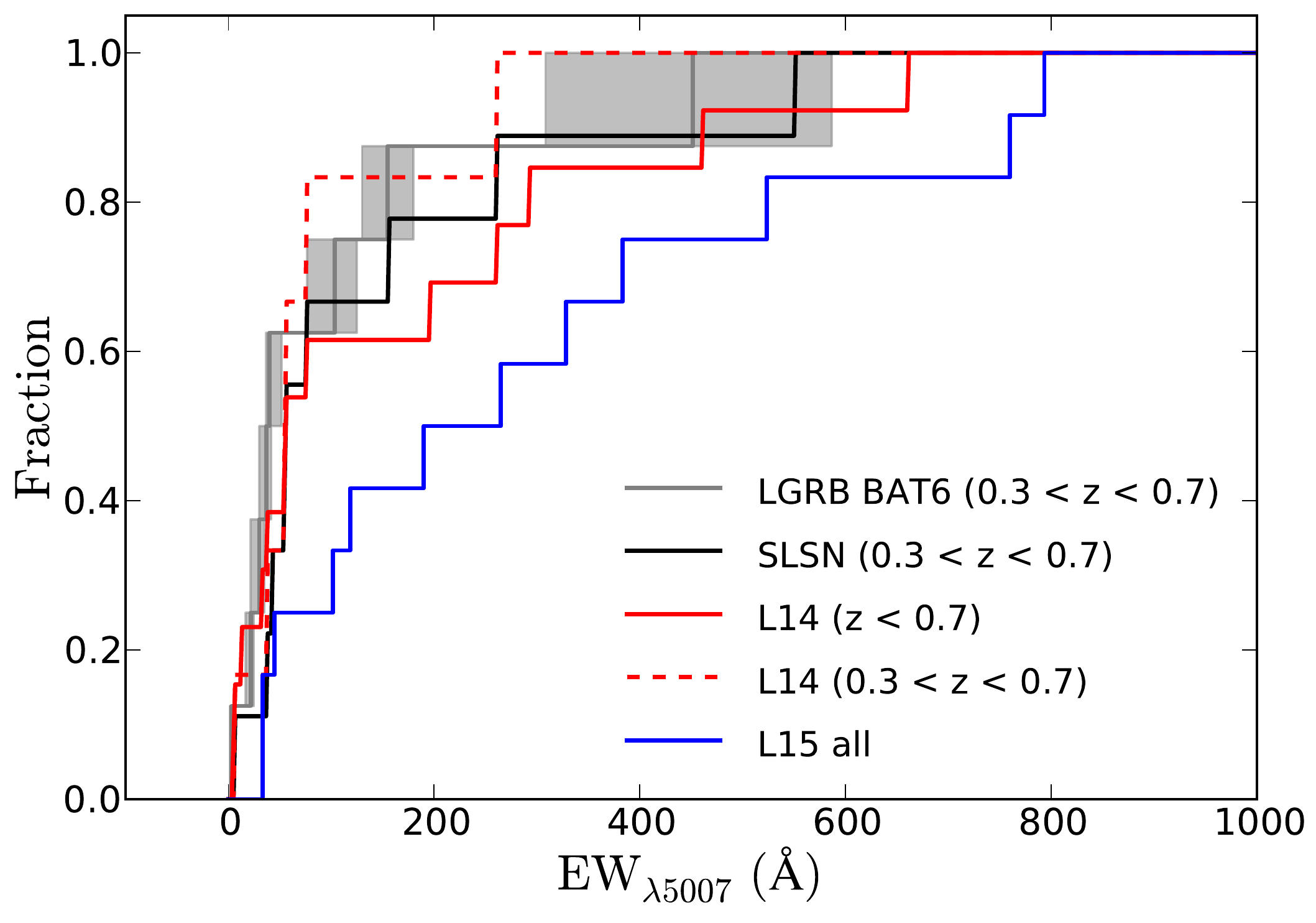}
\caption{Distributions of equivalent widths of the \OIII$\lambda5007$ emission line. The distribution of $0.3 < z < 0.7$ LGRB sample values (grey) is compared to the SLSN sample (Table 1; black) in the same redshift range. Furthermore, we plot distributions of SLSN samples from the literature, namely from L15 (blue) and L14 (red). Errors are not taken into account in the case of SLSN measurements, as they are negligible.}
\label{fig4}
\end{figure}

However, the conclusions of L15 regarding the EW seem to be strongly affected by the domination of low-$z$ SLSN hosts in their sample, compared to the very small fraction of LGRBs at $z<0.3$ of their comparison sample. The effect is illustrated in Figure \ref{fig4}, where we compare the distribution of EW$_{5007}$ of LGRB sample used in this paper\footnote{L15 shows the EW$_{5007}$ distribution built for the TOUGH LGRB complete sample \citep{Hjorth2012}: the distribution is very similar to the one we built with our sample.} to the SLSN distributions presented by L15 and L14. The sample of L15 is built primarily of $z < 0.3$ hosts and is indeed strikingly different from the LGRB distribution. But the sample of L14, including much larger fraction of $z > 0.3$ hosts, agrees much better with the LGRB values. Indeed, taking only the $z > 0.3$ values from the L14 sample, the distribution is essentially the same as the one for LGRB hosts. Finally, we compare the values of the EW$_{5007}$ for the LGRB hosts and SLSN sample defined in Table 1 (e.g. in the $0.3 < z < 0.7$ redshift range; data provided by R. Lunnan and D. Perley, private communication) and confirm that the two distributions are similar. Two SLSN without detected lines (PTF09cwl and PTF09atu) are faint \citep[e.g.][]{Perley2016d}. However, the lack of detected emission lines suggests the lines are not very strong and should therefore not affect our conclusions. Furthermore, using the bootstrap method we confirm that the difference between our sample (high-$z$) and sample from \citet{Leloudas2015} (low-$z$) is not due to small samples used in the comparison.

The previous finding of L15 can therefore mainly be attributed to the apparent redshift evolution of the EW$_{5007}$ in SLSN hosts. According to \citet{Steidel2014} the ionization field, and therefore the strength of the emission lines, in star-forming galaxies is getting stronger with redshift, which is opposite to what we observe in SLSN hosts. A possible explanation is that the observed evolution is due to the size of galaxies in the sky: at lower redshifts the galaxies are resolved and a spectrograph slit can be placed at the positions pointed by SLSN explosions, while at higher redshifts the slit covers an entire galaxy and the obtained spectra represent an average over different nebular regions. This interpretation is supported by two hosts in L15 (hosts of SN1999as and PTF10qaf, the latter being SLSN-II), where spectra taken at different slit positions give different values of EW with the largest at the SLSN explosion cite. Due to the lack of $z < 0.3$ LGRBs in complete samples \citep{Salvaterra2012,Hjorth2012,Perley2015a} we cannot make a more quantitative study of this issue. 
We note that other nebular-line based galaxy characteristics (e.g.\,SFR) are affected by the same redshift-dependent effect. For example, low-$z$ SLSN samples show a strong preference for high sSFR with respect to the field star-forming galaxies \citep{Perley2016d,Chen2016}. It seems that at the redshift range studied in this paper, $0.3 < z < 0.7$, the fraction of SLSN hosts with very high sSFR decreases and is similar to the one observed in the LGRB host samples (Fig. 2). LGRB hosts at this redshift show only a slight preference for high sSFR galaxies with respect to the field star-forming population, though due to low statistics involved in the studies the issue is still unclear \citep{Japelj2016}. In any case, if the spatial resolution is indeed the major cause of the seemingly different properties at low-$z$, it should not have a consequence on the comparison of LGRB and SLSN hosts at $z>0.3$, as their sizes and morphology are similar \citep[e.g.][]{Lunnan2015}.

In the absence of any strong observable difference between LGRB and SLSN-I host galaxy environments we are left with few new clues to improve our understanding of SLSN progenitors. The metallicities observed in SLSN hosts are preferentially low (e.g. see also \citealt{Lunnan2014,Leloudas2015,Perley2016d,Chen2016}), avoiding super solar metallicities. The similar distribution of SLSN and LGRB host galaxy metallicity that we find agrees with the results of \citet{Chen2016}. Indeed, in \citet{Vergani2015} and \citet{Japelj2016} we find a similar preference for $Z < 0.5Z_{\odot}$ for LGRB host galaxies. This is in favour of the progenitor models requiring low metallicities, e.g. the magnetar model.
One should keep in mind that the similarity between LGRB and SLSN host galaxies does not necessarily imply that these two type of events have the same progenitors in general (but see the special case of ultra-long GRB\,111209A and its accompanying SN; \citealt{Greiner2015}).

Finally, given a relatively small number of observed SLSN host galaxies, all current studies are affected by small number statistics. We stress that even though we restricted our analysis to a narrow redshift range, excluding the $z < 0.3$ SLSN population, the size of our sample is still of the same order as previously studied samples. 

\section{Conclusions}

We compared the properties of the host galaxies of SLSNe to those of LGRBs with the aim of looking for similarities and possible differences of the two populations, to try to understand the different results found in the literature on this topic and with the final goal to shed some light on the progenitors of these two phenomena. In contrast to previous studies we used the host galaxies of a complete sample of LGRB as comparison. Furthermore, we performed the comparison in a reduced redshift range, and carefully selected the published stellar masses of SLSN host galaxies.

Our results indicate that, at least within $0.3<z<0.7$, the properties of SLSN and LGRB host galaxies are similar, in terms of their $M_{\it B}$, stellar masses, SFR, sSFR and metallicities. We also show that the host galaxies of SLSN follow the FMR and FPZ relations as field star-forming galaxies. The discrepancies of the results of previous studies are due to the different (biased) comparison samples of LGRB host galaxies used, to the different stellar mass values used among all those studies and to the comparison made over a wide/different redshift range for the two classes of explosions. We do not find evidence of the excess of high-EW SLSN hosts, as found by studies who focused on low-$z$ SLSN samples \citep{Leloudas2015}.

We stress that to improve these results larger samples are needed, e.g. more high-$z$ SLSN and more low-$z$ LGRB host galaxies as well as a complete sample of SLSNe. 


\begin{acknowledgements}
We thank the referee for the helpful comments that improved the paper. We thank R. Lunnan and D. Perley for providing EW measurements for the L14 and P16 samples of SLSN hosts. JJ acknowledges financial contribution from the grant PRIN MIUR 2012 201278X4FL 002 The Intergalactic Medium as a probe of the growth of cosmic structures.
\end{acknowledgements}

\bibliographystyle{aa}
\bibliography{ms_slsn_japelj}

\end{document}